\documentclass[11pt]{article}
\usepackage{moriond2,epsfig}
\usepackage{axodraw}
\usepackage{epsfig}
\usepackage{amsmath}
\usepackage{amssymb}
\usepackage{graphicx}
\usepackage{subfigure}
\usepackage{color}

\input epsf

\begin{document}
\vspace*{4cm}
\title{Understanding the Fierz Ambiguity of Partially Bosonized Theories}
\author{J\"org J\"ackel}

\address{Institut f\"ur Theoretische Physik der Universit\"at Heidelberg\\
  Philosophenweg 16, D-69120 Heidelberg, Germany}

\maketitle

\abstracts{A useful tool in non perturbative studies of fermionic
theories is partial bosonization. However, partial bosonization is
often connected to an ambiguity due to Fierz rearrangement in the
original theory. We discuss two different approximations for the
calculation of the effective action $\Gamma$ with respect to a
spurious dependence on the choice of Fierz transformation: Mean
field theory and the truncated flow of an exact renormalization
group equation for the effective average action.}

\newcommand{\omd}{\Omega _{\rm d}}
\newcommand{\sa}{\sigma _8}
\newcommand{\omdsf}{\bar{\Omega} _{\rm d} ^{\rm sf}}
\newcommand{\omdn}{\Omega _{\rm d}^0}
\newcommand{\omdeq}{\sqrt{1-\bar{\Omega}_{\rm d} (\aeq)}}
\newcommand{\omdroot}{\sqrt{1-\bar{\Omega}_{\rm d} (a)}}
\newcommand{\aeq}{a_{\rm eq}}
\newcommand{\adec}{a_{\rm dec}}
\newcommand{\atr}{a_{\rm tr}}
\newcommand{\wda}{w_{\rm d}}
\newcommand{\wdan}{w_{\rm d}^0}
\newcommand{\kmax}{k_{\rm max}}
\newcommand{\keq}{k_{\rm eq}}
\newcommand{\sla}[1]{\slash\!\!\!#1}
\newcommand{\slad}[1]{\slash\!\!\!\!#1}
\newcommand{\lferm}{\Lambda_{\rm ferm}}
\newcommand{\pmp} {\Phi} 
\newcommand{\pmpc}{\Phi_{\rm cl}} 
\newcommand{\gev}{\textrm{Gev}}
\newcommand{\eloop}{\textrm{1-loop}}
\newcommand{\vv}[2]{V_{\textrm{#1}}^{\textsc{#2}}}
\newcommand{\ww}[2]{W_{\textrm{#1}}^{\textsc{#2}}}
\newcommand{\fpi}{(2\pi)^{-4}}
\newcommand{\tpi}{(2\pi)^{-2}}
\newcommand{\mf}{m_{\rm f}(\Phi_{\rm cl})}
\newcommand{\mfn}{m_{\rm f}^0}
\newcommand{\mfnsq}{\left[\mfn\right]^2}
\newcommand{\mfsq}{\left[\mf\right]^2}
\newcommand{\mplank}{\textrm{M}_{\textrm{P}}}
\newcommand{\fss}[1]{#1\!\!\!/}
\def\frac#1#2{\mathinner{#1\over#2}}

\section{Introduction}
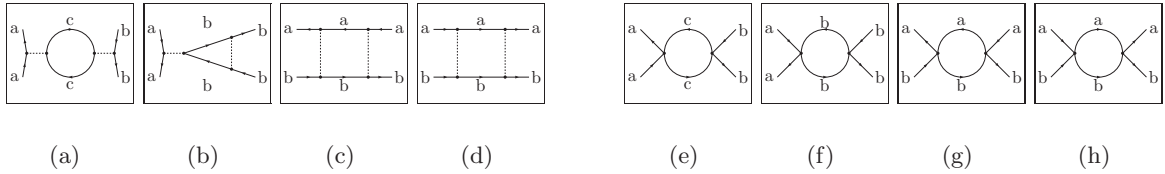
\begin{figure}[t]
\begin{center}
\subfigure[]{\scalebox{0.3}[0.3]{
\fbox{
\begin{picture}(150,120)
\SetOffset(-87,0)
\DashLine(105,60)(130,60){2}
\Vertex(130,60){2}
\ArrowArc(160,60)(-30,0,180)
\ArrowArc(160,60)(-30,180,0)
\Vertex(190,60){2}
\DashLine(190,60)(215,60){2}
\ArrowLine(105,60)(100,90)
\ArrowLine(100,30)(105,60)
\ArrowLine(215,60)(220,30)
\ArrowLine(220,90)(215,60)
\Text(90,90)[]{\scalebox{2.0}[2.0]{a}}
\Text(90,30)[]{\scalebox{2.0}[2.0]{a}}
\Text(160,100)[]{\scalebox{2.0}[2.0]{c}}
\Text(160,20)[]{\scalebox{2.0}[2.0]{c}}
\Text(230,90)[]{\scalebox{2.0}[2.0]{b}}
\Text(230,30)[]{\scalebox{2.0}[2.0]{b}}
\Vertex(105,60){2}
\label{subfig::mass}
\Vertex(215,60){2}
\end{picture}}}}
\subfigure[]{\scalebox{0.3}[0.3]{\fbox{
\begin{picture}(150,120)
\SetOffset(-87,0)
\Vertex(105,60){2}
\ArrowLine(100,90)(105,60)
\ArrowLine(105,60)(100,30)
\DashLine(105,60)(130,60){2}
\Vertex(130,60){2}
\ArrowLine(130,60)(190,80)
\ArrowLine(190,80)(220,90)
\ArrowLine(190,40)(130,60)
\ArrowLine(220,30)(190,40)
\DashLine(190,80)(190,40){2}
\Vertex(190,80){2}
\Vertex(190,40){2}
\Text(90,90)[]{\scalebox{2.0}[2.0]{a}}
\Text(90,30)[]{\scalebox{2.0}[2.0]{a}}
\Text(160,100)[]{\scalebox{2.0}[2.0]{b}}
\Text(160,20)[]{\scalebox{2.0}[2.0]{b}}
\Text(230,90)[]{\scalebox{2.0}[2.0]{b}}
\Text(230,30)[]{\scalebox{2.0}[2.0]{b}}
\label{subfig::vertex}
\end{picture}}}}
\subfigure[]{\scalebox{0.3}[0.3]{\fbox{
\begin{picture}(150,120)
\SetOffset(-87,0)
\ArrowLine(220,90)(190,90)
\ArrowLine(190,90)(130,90)
\ArrowLine(130,90)(100,90)
\ArrowLine(100,30)(130,30)
\ArrowLine(130,30)(190,30)
\ArrowLine(190,30)(220,30)
\DashLine(130,30)(130,90){2}
\DashLine(190,30)(190,90){2}
\Vertex(190,30){2}
\Vertex(130,30){2}
\Vertex(190,90){2}
\Vertex(130,90){2}
\Text(90,90)[]{\scalebox{2.0}[2.0]{a}}
\Text(90,30)[]{\scalebox{2.0}[2.0]{b}}
\Text(160,100)[]{\scalebox{2.0}[2.0]{a}}
\Text(160,20)[]{\scalebox{2.0}[2.0]{b}}
\Text(230,90)[]{\scalebox{2.0}[2.0]{a}}
\Text(230,30)[]{\scalebox{2.0}[2.0]{b}}
\label{subfig::boxa}
\end{picture}}}}
\subfigure[]{\scalebox{0.3}[0.3]{\fbox{
\begin{picture}(150,120)
\SetOffset(-87,0)
\ArrowLine(190,90)(220,90)
\ArrowLine(130,90)(190,90)
\ArrowLine(100,90)(130,90)
\ArrowLine(100,30)(130,30)
\ArrowLine(130,30)(190,30)
\ArrowLine(190,30)(220,30)
\DashLine(130,30)(130,90){2}
\DashLine(190,30)(190,90){2}
\Vertex(190,30){2}
\Vertex(130,30){2}
\Vertex(190,90){2}
\Vertex(130,90){2}
\Text(90,90)[]{\scalebox{2.0}[2.0]{a}}
\Text(90,30)[]{\scalebox{2.0}[2.0]{b}}
\Text(160,100)[]{\scalebox{2.0}[2.0]{a}}
\Text(160,20)[]{\scalebox{2.0}[2.0]{b}}
\Text(230,90)[]{\scalebox{2.0}[2.0]{a}}
\Text(230,30)[]{\scalebox{2.0}[2.0]{b}}
\label{subfig::boxb}
\end{picture}}}}
\hspace{0.8cm}
\subfigure[]{\scalebox{0.3}[0.3]{\fbox{
\begin{picture}(150,120)
\SetOffset(-87,0)
\Text(90,90)[]{\scalebox{2.0}[2.0]{a}}
\Text(90,30)[]{\scalebox{2.0}[2.0]{a}}
\Text(160,100)[]{\scalebox{2.0}[2.0]{c}}
\Text(160,20)[]{\scalebox{2.0}[2.0]{c}}
\Text(230,90)[]{\scalebox{2.0}[2.0]{b}}
\Text(230,30)[]{\scalebox{2.0}[2.0]{b}}
\ArrowLine(130,60)(100,90)
\ArrowLine(100,30)(130,60)
\Vertex(130,60){2}
\ArrowArc(160,60)(30,180,0)
\ArrowArc(160,60)(30,360,180)
\Vertex(190,60){2}
\ArrowLine(220,90)(190,60)
\ArrowLine(190,60)(220,30)
\label{subfig::fermiona}
\end{picture}}}}
\subfigure[]{\scalebox{0.3}[0.3]{\fbox{
\begin{picture}(150,120)
\SetOffset(-87,0)
\Text(90,90)[]{\scalebox{2.0}[2.0]{a}}
\Text(90,30)[]{\scalebox{2.0}[2.0]{a}}
\Text(160,100)[]{\scalebox{2.0}[2.0]{b}}
\Text(160,20)[]{\scalebox{2.0}[2.0]{b}}
\Text(230,90)[]{\scalebox{2.0}[2.0]{b}}
\Text(230,30)[]{\scalebox{2.0}[2.0]{b}}
\ArrowLine(100,90)(130,60)
\ArrowLine(130,60)(100,30)
\Vertex(130,60){2}
\ArrowArc(160,60)(30,180,0)
\ArrowArc(160,60)(30,360,180)
\Vertex(190,60){2}
\ArrowLine(220,90)(190,60)
\ArrowLine(190,60)(220,30)
\label{subfig::fermionb}
\end{picture}}}}
\subfigure[]{\scalebox{0.3}[0.3]{\fbox{
\begin{picture}(150,120)
\SetOffset(-87,0)
\Text(90,90)[]{\scalebox{2.0}[2.0]{a}}
\Text(90,30)[]{\scalebox{2.0}[2.0]{b}}
\Text(160,100)[]{\scalebox{2.0}[2.0]{a}}
\Text(160,20)[]{\scalebox{2.0}[2.0]{b}}
\Text(230,90)[]{\scalebox{2.0}[2.0]{a}}
\Text(230,30)[]{\scalebox{2.0}[2.0]{b}}
\ArrowLine(130,60)(100,90)
\ArrowLine(100,30)(130,60)
\Vertex(130,60){2}
\ArrowArc(160,60)(30,180,0)
\ArrowArc(160,60)(30,360,180)
\Vertex(190,60){2}
\ArrowLine(220,90)(190,60)
\ArrowLine(190,60)(220,30)
\label{subfig::fermionc}
\end{picture}}}}
\subfigure[]{\scalebox{0.3}[0.3]{\fbox{
\begin{picture}(150,120)
\SetOffset(-87,0)
\Text(90,90)[]{\scalebox{2.0}[2.0]{a}}
\Text(90,30)[]{\scalebox{2.0}[2.0]{b}}
\ArrowLine(100,90)(130,60)
\ArrowLine(100,30)(130,60)
\Vertex(130,60){2}
\ArrowArc(160,60)(30,180,0)
\Text(160,20)[]{\scalebox{2.0}[2.0]{b}}
\Text(160,100)[]{\scalebox{2.0}[2.0]{a}}
\ArrowArc(160,60)(-30,180,360)
\Vertex(190,60){2}
\Text(230,90)[]{\scalebox{2.0}[2.0]{a}}
\Text(230,30)[]{\scalebox{2.0}[2.0]{b}}
\ArrowLine(190,60)(220,90)
\ArrowLine(190,60)(220,30)
\label{subfig::fermiond}
\end{picture}}}}
\caption{One loop diagrams for the bosonized model (\ref{subfig::mass}-\ref{subfig::boxb}) and the fermionic
model (\ref{subfig::fermiona}-\ref{subfig::fermionb}). Bosonic lines are dashed, fermionic lines solid and vertices
are marked with a dot. The letters are to visualize in which way the fermionic lines are contracted.
We find a one to one correspondence between the models. Shrinking all bosonic lines to
points we go from \ref{subfig::mass} to \ref{subfig::fermiona}, \ref{subfig::vertex} to \ref{subfig::fermionb} etc..}
\end{center}
\label{fig::summ}
\end{figure}

Mean field theory is a widely used method in many body statistical
physics and quantum field theory, in particular if the ground
state is characterized by condensates and spontaneous symmetry
breaking. For example, mean field solutions of the
Nambu-Jona-Lasinio (NJL) model \cite{Nambu:1961tp} or extensions
of it are one of the main theoretical tools in nuclear physics see
e.g. \cite{Meyer:2000bx,Wetterich:2001sh}. The recent discussion
of color superconductivity at high but realistic baryon density is
mainly based on this method
\cite{Berges:1998rc,Schafer:1999pb,Alford:1999pa,Rapp:2000qa,Oertel:2002pj}.
An example from statistical physics is a mean field description
\cite{Baier:2000yc} of antiferromagnetic and superconducting
condensates in the Hubbard model. Quite generally, mean field
theory (MFT) seems to be well
suited for systems with multifermion interactions and bosonic condensates.\\
Unfortunately, in these systems MFT has a basic ambiguity which is connected with the possibility to perform
Fierz transformations (FT) for the underlying local multifermion interaction. This becomes apparent already in the simplest
NJL-type model (for only one fermion species) with a chirally invariant pointlike four fermion interaction:
\begin{eqnarray}
\label{equ::faction}
\Gamma_{k}=\int d^{4}x
&[&\bar{\psi}i\fss{\partial}\psi
+\frac{1}{2}\lambda_{\sigma,k}[(\bar{\psi}\psi)^{2}-(\bar{\psi}\gamma^{5}\psi)^2]
+\frac{1}{2}\lambda_{V,k}[(\bar{\psi}\gamma^{\mu}\psi)^2]
+\frac{1}{2}\lambda_{A,k}[(\bar{\psi}\gamma^{\mu}\gamma^{5}\psi)^{2}].
\end{eqnarray}
Due to the Fierz identity
\begin{equation}
\label{equ::fierz}
\left [(\bar{\psi}\gamma^{\mu}\psi)^{2}-(\bar{\psi}
\gamma^{\mu}\gamma^{5}\psi)^{2}\right]
+2\left[(\bar{\psi}\psi)^{2}-(\bar{\psi}\gamma^{5}\psi)^{2}\right ]=0
\end{equation}
only two of the quartic couplings are independent and we write
\begin{equation}
\label{equ::invariant}
\lambda_{\sigma}=\bar{\lambda}_{\sigma}-2\gamma\bar{\lambda}_{V},
\quad \lambda_{V}=(1-\gamma)\bar{\lambda}_{V},\quad \lambda_{A}=\gamma\bar{\lambda}_{V}.
\end{equation}
The parameter $\gamma$ is redundant since it multiplies just the vanishing expression \eqref{equ::fierz}.
No physical quantity can depend on $\gamma$ in a full computation of the functional integral for
partition function and expectation values of field operators.\\
The index $k$ for the effective action denotes a cutoff scale. We will see in sect.
\ref{sec::mean} that the MFT results do strongly depend on $\gamma$, limiting their quantitative
reliability. For example the critical coupling for the onset of a non vanishing condensate
$\sigma\sim\langle\bar{\psi}(1-\gamma^{5})\psi\rangle$ depends strongly on $\gamma$
for fixed physical couplings $\bar{\lambda}_{\sigma}$ and $\bar{\lambda}_{V}$.
MFT is tightly connected to the method of partial bosonization. Indeed, MFT can be thought of as simply
performing the fermionic functional integral of the partially bosonized model introduced below.
In order to make progress one has to find a method where MFT appears as some type of first step in a more
systematic expansion. As a test of such a method one may investigate if the results become independent of
$\gamma$ as it should be. In this talk (for a more detailed discussion see \cite{me})
we want to discuss such a method based on the
exact renormalization group (RG) equation for the effective
average action \cite{Wetterich:1993be}.\\
Perturbation theory is an alternative approach to the model \eqref{equ::faction} (see Fig. \ref{fig::summ}).
Results of perturbation theory are unambiguous. However, perturbation theory is
limited to small coupling. Therefore we cannot observe the interesting phenomena
of spontaneous symmetry breaking (SSB). Doing RG-improvement it is
possible to observe the onset of SSB \cite{Aoki:1997fh}. Nevertheless, in the simple model of \eqref{equ::faction}
it is impossible to proceed to the SSB phase. To do this we would need to enlarge the truncation to
include eight and higher fermion interactions which is quite difficult.\\
Partial bosonization seems to be the ideal remedy to this difficulty
\cite{Klevansky:1992qe,Ellwanger:1994wy,Alkofer:1996ph,Berges:1999eu,Jungnickel:1997ke}.
Using this technique the model \eqref{equ::faction} can be rewritten as an equivalent Yukawa type
model with scalars $\phi$, vectors $V^{\mu}$ and axial vectors $A^{\mu}$ representing
the corresponding fermion bilinears:
\begin{eqnarray}
\label{equ::baction}
\mathcal{L}_{B}=\int d^{4}x&\{&\mu^{2}_{\sigma}\phi^{\star}\phi
+\frac{1}{2}\mu^{2}_{V}(V^{\mu})^{2}+\frac{1}{2}\mu^{2}_{A}(A^{\mu})^{2}
\\\nonumber
&+&h_{\sigma}\left[ \bar{\psi}\left(\frac{1+\gamma^{5}}{2}\right)\phi\psi
-\bar{\psi}\left(\frac{1-\gamma^{5}}{2}\right)\phi^{\star}\psi\right]
+h_{V}\bar{\psi}\gamma^{\mu}V^{\mu}\psi
+h_{A}\bar{\psi}\gamma^{\mu}\gamma^{5}A^{\mu}\psi
\end{eqnarray}
Taking
\begin{equation}
\label{equ::bosocouplings}
\mu^{2}_{\sigma}=\frac{h^{2}_{\sigma}}{2\lambda_{\sigma}},
\quad\mu^{2}_{V}=-\frac{h^{2}_{V}}{\lambda_{V}},
\quad\mu^{2}_{A}=-\frac{h^{2}_{A}}{\lambda_{A}}
\end{equation}
at some
cutoff scale $k=\Lambda$ this model is equivalent to the NJL-type model \eqref{equ::faction}. Indeed,
partial bosonization is nothing more than the introduction of a factor of unity into the functional integral.\\
Spontaneous symmetry breaking can now be dealt with by computing
the effective potential for $\phi$ and looking for a minimum at $\phi\neq 0$. For example, a term $\sim \phi^{4}$
stands for an eight quark interaction. Unfortunately, partial bosonization brings back the ''Fierz ambiguity''
of MFT.\\
\section{Critical couplings from MFT} \label{sec::mean}
For a mean field calculation we treat the fermionic fluctuations in a homogenous background of
a $\phi\sim\bar{\psi}(1-\gamma^{5})\psi$ field.
Partial bosonization introduces just such a composite field and we use the action \eqref{equ::baction}.
MFT means that we perform the functional integral in a homogeneous background $\phi$-field.
For the correction to the mass of the $\phi$-field the corresponding Feynman diagram is given in Fig. \ref{subfig::mass}.
Since we only want to determine the critical couplings we are
satisfied to calculate the mass term $\sim \phi^{\star}\phi$
and look when it turns negative. Therefore we do not need to consider the other background fields
such as $V^{\mu}\sim\bar{\psi}\gamma^{\mu}\psi$ and $A^{\mu}\sim\bar{\psi}\gamma^{\mu}\gamma^{5}\psi$. So,
for these fields we set $V^{\mu}=A^{\mu}=0$.
Including the fluctuations from $k=\Lambda$ to $k=0$ we find:
\begin{equation}
\Gamma=\Gamma_{\Lambda}+\Delta\Gamma^{\textrm{MFT}}
=\left(\mu^{2}_{\sigma,\Lambda}-\frac{1}{8\pi^{2}}h^{2}_{\sigma,\Lambda}\Lambda^{2}\right)\phi^{\star}\phi
+const+{\mathcal{O}}\left((\phi^{\star}\phi)^{2}\right),
\end{equation}
where we have expanded in powers of the $\phi$-field to better see the mass term.\\
The mass term turns negative at the critical coupling
\begin{equation}
\label{equ::crit}
\lambda^{\textrm{crit}}_{\sigma,\Lambda}
=\frac{h^{2}_{\sigma,\Lambda}}{2\mu^{2}_{\sigma,\Lambda}}=\frac{4\pi^{2}}{\Lambda^{2}},\quad
\bar{\lambda}^{\textrm{crit}}_{\sigma,\Lambda}=\frac{4\pi^{2}}{\Lambda^{2}}
+2\gamma\bar{\lambda}_{V,\Lambda}
\end{equation}
Where we used Eqs. \eqref{equ::bosocouplings} and \eqref{equ::invariant}
to express the result in terms of the underlying fermionic model. The result depends
on the unphysical parameter $\gamma$ and is therefore ambiguous.
\section{Invariant Bosonic Flow} \label{sec::redef}
In the context of the RG-equation for the effective average action the way to
specify an approximation is to choose an ansatz for the effective average action. Using
\eqref{equ::baction} as an ansatz it is possible to calculate the flow equations for
the (now $k$-dependent) couplings and mass terms. This calculation includes not only the mass shift
Fig. \ref{subfig::mass} but also the vertex corrections \ref{subfig::vertex}. However the
results (e.g. the critical coupling) still depend strongly on $\gamma$ and therefore we do
not reproduce perturbation theory.\\
At this point we note a discrepancy between the claim that bosonization is an exact identity
and the fact that we do \emph{not} reproduce perturbation theory. Furthermore we get a dependence on the
unphysical Fierz parameter $\gamma$ (actually this is an expression of the fact that we do not
reproduce perturbation theory which is invariant).\\
Of course, the exact equivalence between the bosonized and the non bosonized model is ensured only
if we calculate the complete flow. When we do approximations it might be violated. That is what has
happened in our model. Where did this happen? The bosonization procedure
cancelled all four fermion interactions at the bosonization scale $\Lambda$. However, during the
flow to $k<\Lambda$ new four fermion interactions are generated by box diagrams with internal
bosonic lines Figs. \ref{subfig::boxa} and \ref{subfig::boxb}. These are not included in the
truncation to \eqref{equ::faction}. Nevertheless, all diagrams of Figs. \ref{subfig::mass}-\ref{subfig::boxb} are
of the same order $\sim h^{4}$. So it seems inconsistent to neglect the two box diagrams and
therefore the generated four fermion interactions. However, including a four fermion
interaction into the truncation seems not to be a sensible thing to do since we bosonized to get rid
of these complicated multi fermion interactions. Luckily,
a method to absorb four fermion interactions into the bosonic flow has been developed in in \cite{Gies:2002nw}.
Applying this method to the model \eqref{equ::baction} we can absorb the interactions generated
by the diagrams Figs. \ref{subfig::boxa} and \ref{subfig::boxb}. This completes the flow in the sense that
now all diagrams at this order are taken into account. We are now able to reproduce
one loop perturbation theory in the bosonic model. A thorough analysis\cite{me} shows that the RG calculation
for this truncation is now indeed equivalent to the RG calculation for the purely fermionic
model specified by \eqref{equ::faction}. Since the latter is invariant under FT's for
the initial action the former also exhibits this feature.
\section{Conclusions}
We have shown that MFT leads to ambiguous results depending on the choice of FT
for the underlying fermionic theory. An RG calculation which includes also
the vertex correction for the Yukawa coupling does not improve on this point.\\
The reason for this problem is that we have neglected the four fermion interactions generated during the flow
even though they are of the same order in the coupling. Including these interactions we are able
to reproduce (RG-improved) one-loop perturbation theory. Moreover we find invariance of the result under
FT's.\\
Having established a way to produce FT invariant results at the lowest level it seems now possible to look
toward more complicated (and more useful) truncations e.g. including kinetic terms for the bosons.
It might be difficult to achieve complete independence of the results on the FT but one can hope
at least for a weaker dependence.
\section*{Acknowledgments}
The author would like to thank Tobias Baier,  J\"urgen Berges, Eike Bick, Michael Doran, Kai Schwenzer and
Christof Wetterich for helpful discussions and the organizers for a fruitful conference.

\end{document}